\documentclass{JHEP3} 
\usepackage{amsmath,amssymb,bbm}

\newcommand{\ee}{\mathrm{e}} 
\newcommand{\im}{\mathrm{i}} 
\newcommand{\spc}{{\,}} 
\newcommand{\ft}[2]{{\textstyle\frac{#1}{#2}}}

\newcommand{\Xh}{\hat{X}}

\newcommand{\abs}[1]{\lvert #1\rvert} 
 
  \DeclareMathOperator{\sgn}{sgn} 
\newcommand{\pd}{\partial}

\newsavebox{\uuunit} \sbox{\uuunit} {\setlength{\unitlength}{0.825em} 
\begin{picture}
	(0.6,0.7) \thinlines \put(0,0){\line(1,0){0.5}} \put(0.15,0){\line(0,1){0.7}} \put(0.35,0){\line(0,1){0.8}} \multiput(0.3,0.8)(-0.04,-0.02){10}{\rule{0.5pt}{0.5pt}} 
\end {picture}}

\title{First-order flow equations for extremal black holes \\
in very special geometry}

\author{G.~L.~Cardoso,$^{a}$ A.~Ceresole,$^{b}$ G.~Dall'Agata,$^{c}$ J.~M.~
Oberreuter$^a$ and J.~Perz$^{a,d}$\\
$^a$ Arnold Sommerfeld Center for Theoretical Physics\\
Department f\"ur Physik, Ludwig-Maximilians-Universit\"at M\"unchen \\
Theresienstr. 
37, 80333 M\"unchen, Germany \\
$^b$ INFN, Sezione di Torino \& Dipartimento di Fisica Teorica Universit\`a di Torino \\
Via Pietro Giuria 1, 10125 Torino, Italy \\
$^c$ Dipartimento di Fisica ``Galileo Galilei'' \& INFN, Sezione di Padova\\
Universit\`a di Padova, Via Marzolo 8, 35131 Padova, Italy \\
$^d$ Max-Planck-Institut f\"ur Physik\\
F\"ohringer Ring 6, 80805 M\"unchen, Germany }

\abstract{ We construct interpolating solutions describing single-center static extremal non-supersymmetric black holes in four-dimensional $N=2$ supergravity theories with cubic prepotentials. 
To this end, we derive and solve first-order flow equations for rotating electrically charged extremal black holes in a Taub-NUT geometry in five dimensions. 
We then use the connection between five- and four-dimensional extremal black holes to obtain four-dimensional flow equations and we give the corresponding solutions.

}

\preprint{ DFPD-07TH10 \\
LMU-ASC 38/07 \\
MPP-2007-73 }

\begin{document}

\section{Introduction} \setcounter{equation}{0}

The radial dependence of single-center static supersymmetric black hole solutions \cite{ Ferrara:1995ih,Strominger:1996kf,Ferrara:1996dd,Ferrara:1996um} in four-dimensional $N=2$ supergravity theories at the two-derivative level can be determined by solving a set of first-order differential (flow) equations \cite{Ferrara:1997tw,Moore:1998pn,Denef:2000nb}. 
Such solutions are given in terms of harmonic functions \cite{Ferrara:1995ih,Ferrara:1996dd,Behrndt:1996jn, Sabra:1997kq,Sabra:1997dh,Behrndt:1997ny}. 
Furthermore, these black holes can be connected to supersymmetric black hole solutions \cite{Sabra:1997yd,Chamseddine:1998yv,Gauntlett:1998fz,Gauntlett:2004qy} in five-dimensional $N=2$ supergravity theories. 
This link, which more generally holds for extremal black holes, is implemented \cite{Gaiotto:2005gf,Gaiotto:2005xt} by placing the five-dimensional black hole in a Taub-NUT geometry, and by using the modulus of the Taub-NUT space to change between the five- and the four-dimensional description. 
The 5d/4d-connection thus provides a unified treatment of extremal black holes arising in five- and four-dimensional $N=2$ supergravity theories with cubic prepotentials. 
In this note, we focus on single-center extremal non-supersymmetric black holes \cite{Ferrara:1997tw,Gibbons:1997cc,Goldstein:2005hq, Tripathy:2005qp,Kallosh:2006bt,Kallosh:2006ib} in such theories and we use the description based on very special geometry in five dimensions, which is simpler to deal with than its four-dimensional counterpart, to construct interpolating black hole solutions in four dimensions (i.e.~solutions with full radial dependence which interpolate between Minkowski spacetime at spatial infinity and the near-horizon 
geometry $AdS_2 \times S^2$).

The features of supersymmetric black holes mentioned above continue to hold for certain classes of extremal non-supersymmetric black holes. 
For instance, it was shown in \cite{Kallosh:2006ib} that the so-called STU-model admits interpolating black hole solutions of this kind which are constructed out of harmonic functions. 
In \cite{Ceresole:2007wx} first-order flow equations were found for certain extremal non-supersymmetric black holes in four dimensions. 
A generalization of some of the findings of \cite{Ceresole:2007wx} was given in \cite{Andrianopoli:2007gt}. 

First-order flow equations exist when the effective black hole potential can be expressed in terms of a ``superpotential'' $W$. 
Then, the effective two-derivative Lagrangian can be written as a sum of squares of first-order flow equations involving $W$. 
The rewriting of the black hole potential in terms of $W$ is, however, not unique \cite{Ceresole:2007wx}. 
A given black hole potential may thus give rise to different first-order flow equations, and the resulting black hole solutions may or may not be supersymmetric.

In this note, we use very special geometry to construct first-order flow equations for five-dimensional rotating electrically charged extremal black holes in a Taub-NUT geometry, generalizing the results obtained in \cite{Larsen:2006xm} for static extremal black holes in asymptotically flat spacetime in five dimensions. 
The ``superpotentials'' $W_5$ we employ are constructed out of the five-dimensional central charge by rotating the electric charges with non-trivial elements belonging to the invariance group of the inverse matrix $G^{AB}$ associated with the kinetic terms for the Maxwell fields. 
We solve the flow equations and obtain interpolating solutions describing extremal black holes in a Taub-NUT geometry in five dimensions. 
Then we use the 5d/4d-connection to obtain four-dimensional first-order flow equations, based on four-dimensional ``superpotentials'' $W_4$, from the five-dimensional flow equations. 
In this way, we give a new interpretation of the results of \cite{Ceresole:2007wx}. 
The solutions to the four-dimensional flow equations that we present describe single-center static dyonic extremal black holes in four dimensions, whose magnetic charge is the NUT charge. 
Some of the non-supersymmetric solutions in four dimensions we find are connected to supersymmetric solutions in five dimensions, as already observed in \cite{Cardoso:2007rg}. 
This feature is related to the $U(1)$-fibration of the Taub-NUT geometry \cite{Nilsson:1984bj,Duff:1997qz}. 
For this set of solutions, the associated first-order flow equations in four dimensions may hence be explained in terms of hidden supersymmetry. 
We refer to the last section for a summary of our results.

\section{Extremal black holes in five and four dimensions \label{secdic} } \setcounter{equation}{0}

Extremal black holes in five dimensions can be related to extremal black holes in four dimensions by placing the five-dimensional black hole in a Taub-NUT geometry. 
In the vicinity of the NUT charge, spacetime looks five-dimensional, whereas far away from it spacetime looks four-dimensional. 
This connection was first established in \cite{Gaiotto:2005gf,Gaiotto:2005xt} for supersymmetric black holes in the context of $N=2$ supergravity theories that in four dimensions are based on cubic prepotentials, and was further discussed in \cite{Behrndt:2005he}. 

Here, we focus on single-center rotating extremal black holes in five dimensions which are connected to single-center static extremal black holes in four dimensions. 
We do this in the context of $N=2$ supergravity theories with cubic prepotentials (cf.~appendix \ref{sec:VSG} for the conventions used). 
In five dimensions, the black holes we consider are electrically charged and may carry one independent angular momentum parameter. 
The NUT charge is denoted by $p^0$ and is positive. 
In four dimensions, the associated static black holes have charges $(p^0, q_0, q_A)$, where $q_0$ is related to the rotation in five dimensions \cite{Gaiotto:2005gf,Gaiotto:2005xt}.

The corresponding line elements are related by dimensional reduction over a compact direction of radius $R$. 
In the context of five-dimensional theories based on $n$ abelian gauge fields $A_5^A$ and real scalar fields $X^A$ ($A= 1, \dots, n$) coupled to gravity, the reduction is based on the following standard formulae (see for instance \cite{Gunaydin:1983bi}), 
\begin{eqnarray}
	ds^2_5 &=& \ee^{2 \phi} \, ds^2_4 + \ee^{- 4 \phi} \,(d x^5 - A^0_4)^2 \;\;\;,\;\; dx^5 = R \, d \psi \;, \nonumber\\
	A_5^A &=& A_4^A + C^A \, (d x^5 - A_4^0) \;,\nonumber\\
	{\hat X}^A &=& \ee^{- 2 \phi} \, X^A\;, \label{dic} 
\end{eqnarray}
where the $A_4^I$ denote the four-dimensional abelian gauge fields (with $I = 0, A$). 
The rescaled scalar fields ${\hat X}^A$ and the Kaluza-Klein scalars $C^A$ are combined into the four-dimensional complex scalar fields $z^A$ \cite{Gunaydin:1983bi}, 
\begin{eqnarray}
	z^A = C^A + \im {\hat X}^A \;. 
	\label{zA} 
\end{eqnarray}
The five-dimensional scalar fields $X^A$ satisfy the constraint \eqref{constrV}. 
The quantity ${\rm e}^{- 6 \phi}$ is related to the four-dimensional K\"ahler potential $K(z, {\bar z})$ given in \eqref{emK} by 
\begin{eqnarray}
	{\rm e}^{- 6 \phi} = \frac{1}{6 V} \, C_{ABC} \, {\hat X}^A {\hat X}^B {\hat X}^C = \frac{1}{8V} \, {\rm e}^{-K} \;. 
	\label{phiK} 
\end{eqnarray}

We take the five-dimensional line element and the five-dimensional gauge fields $A_5^A$ to be given by 
\begin{eqnarray}
	\label{metric} ds_5^2 &=& G_{MN} \, dx^M \, dx^N = - f^2 (r) \, (dt + w)^2 + f^{-1}(r) \, ds^2_{\rm HK} \;, \\
	\label{AA} A^A_5 &=& \chi^A (r) \, (dt + w) \;, 
\end{eqnarray}
where $ds^2_{\rm HK}$ describes the line element of a four-dimensional hyper-K\"ahler manifold. 
We set 
\begin{eqnarray}
	ds^2_{\rm HK} &=& N \, \left(dr^2 + r^2 (d \theta^2 + \sin^2 \theta \, d \varphi^2) \right) + R^2 N^{-1} ( d\psi + p^0 \, \cos \theta \, d\varphi )^2 \;, \nonumber\\
	w &=& w_5 (r) \, ( d\psi + p^0 \, \cos \theta \, d \varphi) + w_4 (r) \, \cos \theta \, d \varphi \;. 
	\label{tnhyper} 
\end{eqnarray}
Here $\theta \in [0, \pi],\, \varphi \in [0, 2 \pi), \, \psi \in [0, 4 \pi)$ and $N$ denotes a harmonic function in three spatial dimensions, 
\begin{equation}
	\label{N} N = h^0 + \frac{p^0 \, R }{r} \;\;\;, \;\;\; p^0 > 0 \;. 
\end{equation}
When $h^0 =0$ and $p^0 =1$, the line element $ds^2_{\rm HK}$ describes a four-dimensional flat space, whereas, when $h^0 > 0$, it describes a Taub-NUT space. 
In the following, we will take $h^0> 0 \,,\, p^0 > 0$.

Using \eqref{dic} and reducing over $\psi$ results in 
\begin{equation}
	\ee^{-4\phi} = \frac{1}{fN} - \left(\frac{f w_5}{R}\right)^2 \label{phi4} 
\end{equation}
as well as 
\begin{equation}
	A_4^0 = \ee^{4\phi}\frac{f^2 w_5}{R}(dt + w_4\cos\theta\,d\varphi) - R p^0 \cos\theta\,d\varphi \spc. 
\end{equation}
The associated four-dimensional line element and the four-dimensional gauge fields $A_4^A$ are 
\begin{eqnarray}
	\label{line4} ds_4^2 &=& -\ee^{2U}(dt + w_4\cos\theta\,d\varphi)^2 + \ee^{-2U} \left(dr^2 + r^2 (d \theta^2 + \sin^2\theta\,d\varphi^2) \right) \;, \\
	\label{gauge4} A_4^A &=& \frac{\ee^{4\phi}}{fN} \, \chi^A \, (dt + w_4\cos\theta\,d\varphi) \spc, 
\end{eqnarray}
with 
\begin{equation}
	\ee^{2U} = \ee^{2\phi}\frac{f}{N} \;. 
	\label{rel2u2phi} 
\end{equation}
The Kaluza-Klein scalars $C^A$ are given by 
\begin{equation}
	C^A = \frac{w_5}{R} \, \chi^A \spc. 
	\label{casol} 
\end{equation}
For a static black hole in four dimensions $w_4 =0$. 
The resulting black hole carries charges $(p^0, q_0, q_A)$, but no magnetic charges $p^A$.

In the following, we will set the value of the Taub-NUT modulus to $R=1$, for convenience.

\section{Flow equations in five dimensions \label{sec:flow5dw5}}

Here we derive first-order flow equations for five-dimensional rotating electrically charged black holes in the geometry \eqref{tnhyper}.

The bosonic part of the five-dimensional $N=2$ supergravity action $S$ reads 
\begin{eqnarray}
	\frac{8 \pi G_5}{V} \; S &=& \int d^5 x \sqrt{-G} \left( {\cal R} - g_{ij} \, \partial_M \phi^i \, \partial^M \phi^j - \frac12 \, G_{AB} F^A_{MN} \, F^{B MN} \right) \nonumber\\
	&& - \frac{1}{6V} \int C_{ABC} \, F^A \wedge F^B \wedge A^C \;. 
	\label{action5} 
\end{eqnarray}
We use the conventions of \cite{Cardoso:2007rg}, which we summarize in appendix \ref{sec:VSG}. 
In particular, we set $2 V = 1$.

We take $\phi^i = \phi^i (r)$ and insert the ansatz \eqref{metric} and \eqref{AA} into the action \eqref{action5}. 
We write the result as 
\begin{equation}
	\frac{8 \pi G_5}{V} \; S = S_1 + S_2 \;, 
\end{equation}
where $S_2$ contains only terms that are proportional to $w$ and derivatives thereof. 
We obtain (we refer to appendix \ref{bulkaction5} for some of the details) 
\begin{eqnarray}
	S_1 &=& \frac12 \int dt \, dr \, d \theta \, d \varphi \, d \psi\, \sin \theta \nonumber\\
	&& \hskip 2cm \left[ -3 r^2 f^{-2} (f')^2 - 2 r^2 g_{ij} {\phi'^i} {\phi'^j} + 2 r^2 f^{-2} G_{AB} {\chi'^A} \, {\chi'^B} + 2 \partial_r \left(r^2 f^{-1} f' \right) \right]\;, \nonumber\\
	S_2 &=& \frac12 \int dt \, dr \, d \theta \, d \varphi \, d \psi\, \sin \theta \nonumber\\
	&& \left[ \frac{f}{r^2 N}\left[ (p^0 w_5 + w_4)^2 + r^4 N^2 w_5'^2 + r^2 \cot^2\theta\,w_4'^2\right] \left(f^2 - 2 G_{AB}\chi^A \chi^B\right) \right. 
	\nonumber\\
	&&\left. 
	+ \frac{2}{3V} C_{ABC} \, (p^0 w_5 + w_4) \, \chi^A \chi^B \chi^C \, w_5' \right]\;, \label{bulkeva} 
\end{eqnarray}
where $\; '= \partial_r \;$. 

The terms in $S_1$ can be written as 
\begin{eqnarray}
	S_1 &=& \frac12 \int dt \, dr \, d \theta \, d \varphi \, d \psi\, \sin \theta \left[-3 r^2 f^{-2} (f')^2 - 2 r^2 g_{ij} {\phi'^i} {\phi'^j} \right. 
	\nonumber\\
	&& \qquad \left. 
	+ 2 r^{-2} f^{-2} G_{AB} \left( r^2 {\chi'^A} + f^2 G^{AC} q_C \right) \left( r^2 {\chi'^B} + f^2 G^{BD} q_D \right) - 2 r^{-2} f^2 \,q_A G^{AB} q_B \right. 
	\nonumber\\
	&& \left. 
	\qquad + 2 \partial_r \left(r^2 f^{-1} f' -2 q_A \, \chi^A \right) \right] \;. 
	\label{action1rew} 
\end{eqnarray}
The term proportional to $q_A G^{AB} q_B$ is the so-called black hole potential \cite{Ferrara:1996dd,Ferrara:2006xx} which, with the help of \eqref{ginvdXdX}, can be written as 
\begin{equation}
	q_A G^{AB} q_B = \frac23 |Z_5|^2 + g^{ij} \,\partial_i |Z_5| \, \partial_j |Z_5| \;, \label{effpotz} 
\end{equation}
where 
\begin{equation}
	Z_5 = q_A \, X^A 
\end{equation}
denotes the (real) central charge in five dimensions. 
The rewriting \eqref{effpotz} is, however, not unique, as discussed in \cite{Ceresole:2007wx} in the four-dimensional context. 
Whenever the inverse vector fields kinetic matrix $G^{AB}$ possesses an invariance group with elements $R^A{}_B$, i.e. 
\begin{equation}
	R^A{}_C \, G^{CD} \, R^B{}_D = G^{AB}, \label{condition} 
\end{equation}
and if $R^A{}_B$ is a constant real matrix, then $q_A G^{AB} q_B$ can more generally be written as\footnote{There may exist other rewritings of the black hole potential which are not captured by \eqref{w5}.} 
\begin{equation}
	q_A G^{AB} q_B = \frac23 |W_5|^2 + g^{ij} \,\partial_i |W_5| \, \partial_j |W_5| \;, \label{effpot} 
\end{equation}
where 
\begin{equation}
	W_5 = Q_A \, X^A \;\;\;,\;\;\; Q_A = q_B \, R^B{}_A \;. 
	\label{w5} 
\end{equation}
The case \eqref{effpotz} is contained in \eqref{effpot} with $R^A{}_B = \delta^A{}_B$. 
Specific examples based on \eqref{w5} will be discussed in section \ref{sec:W5V}.

The rewriting of the black hole potential in terms of \eqref{w5} results in a rewriting of $S_1$, as follows. 
First, we observe that $S_1$ can be cast in the form \eqref{action1rew}, with $q_A$ replaced by $Q_A$ in the black hole potential. 
Then, using (\ref{effpot}) we obtain 
\begin{eqnarray}
	\label{stationarySBPS} S_1 &=& \frac12 \int dt \, dr \, d \theta \, d \varphi \, d \psi\, \sin \theta \biggl[-3 \tau^2 f^2 \left( \partial_{\tau} f^{-1} - \frac23 |W_5| \right)^2 \nonumber\\
	&& - 2 \tau^2 g_{ij} \left( \partial_{\tau} \phi^i + f g^{il} \partial_l |W_5| \right) \left( \partial_{\tau} \phi^j + f g^{jk} \partial_k |W_5| \right) \nonumber\\
	&& \left. 
	+ 2 \tau^2 f^{-2} G_{AB} \left(\partial_{\tau} {\chi^A} - f^2 G^{AC} q_C \right) \left( \partial_{\tau} {\chi^B} - f^2 G^{BD} q_D \right) \right. 
	\nonumber\\
	&& + 2 \partial_r \left(r^2 f^{-1} f' -2 q_A \, \chi^A - 2 f \, |W_5| \right) \biggr] \;, 
\end{eqnarray}
where 
\begin{equation}
	\tau = \frac{1}{r} \;. 
\end{equation}
The last term in \eqref{stationarySBPS} denotes a total derivative, and we will comment on its interpretation below. 
Thus, up to a total derivative term, $S_1$ is expressed in terms of squares of first-order flow equations which, when requiring stationarity of $S_1$ with respect to variations of the fields, result in 
\begin{subequations}
	\label{BPSflow5d} 
	\begin{align}
		\partial_{\tau} f^{-1} &= \frac23 |W_5| \;, \label{flowfi} \\
		\partial_{\tau} {\chi^A} &= f^2 G^{AB} q_B \;, \label{flowchi} \\
		\partial_{\tau} \phi^i &= - f g^{ij} \partial_j |W_5| \;. 
		\label{flowphi} 
	\end{align}
\end{subequations}
They are analogous to the flow equations for static supersymmetric black holes in asymptotically flat spacetime in five dimensions derived in \cite{Larsen:2006xm}.

Next, we rewrite $S_2$ as a sum of squares, as follows. 
Defining 
\begin{equation}
	\label{psitilde} \tilde\chi^A = \chi^A + s f \,\tilde X^A \;\;\;,\;\;\; \tilde X^A = R^A{}_B \, X^B \spc, 
\end{equation}
with a certain proportionality constant $s$, to be determined in \eqref{s}, we rewrite 
\begin{equation}
	G_{AB}\chi^A\chi^B = G_{AB}\tilde\chi^A\tilde\chi^B - 2 s f \, G_{AB} \, \tilde\chi^A \, \tilde X^B + \frac32 s^2 f^2 
\end{equation}
where we used $G_{AB} \, R^A{}_C \, R^B{}_D = G_{CD}$, and 
\begin{equation}
	C_{ABC}\chi^A\chi^B\chi^C = C_{ABC} \left[\tilde\chi^A\tilde\chi^B(\tilde\chi^C - 3sf \, \tilde X^C) + 3 s^2 f^2 \tilde\chi^A \, \tilde X^B \, \tilde X^C - s^3 f^3 \tilde X^A \tilde X^B \tilde X^C \right] \;. 
\end{equation}
Then, we obtain for $S_2$, 
\begin{eqnarray}
	S_2 &=& \frac12 \int dt \, dr \, d \theta \, d \varphi \, d \psi\, \sin \theta \nonumber\\
	&&\left\{ -\frac{2f}{r^2 N}\left[ (p^0 w_5 + w_4)^2 + r^4 N^2 w_5'^2 + r^2 \cot^2\theta\,w_4'^2\right] G_{AB}\tilde\chi^A\tilde\chi^B \right. 
	\nonumber\\
	&&+\frac{2}{3V} (p^0 w_5 + w_4) w_5' \, C_{ABC} \, \tilde\chi^A\tilde\chi^B(\tilde\chi^C - 3sf \tilde X^C)\nonumber\\
	&&+\frac{f^2}{ r^2 N}\left[ (p^0 w_5 + w_4)^2 + r^4 N^2 w_5'^2 + r^2 \cot^2\theta\,w_4'^2\right] \left[f ( 1 - 3 s^2 )+ 4 s \tilde\chi^A G_{AB} \tilde X^B \right]
	\nonumber\\
	&& \left. 
	+\frac{2s^2 f^2}{3 V} (p^0 w_5 + w_4) \,w_5' \,C_{ABC} \left(3 \tilde\chi^A \tilde X^B \tilde X^C- s f \tilde X^A \tilde X^B \tilde X^C \right) \right\} \spc. 
\end{eqnarray}
The first two lines already form a sum of squares (of $\tilde \chi^A$). 
The last two can also be combined into perfect squares provided that $s = \pm 1$ and that $R^A{}_B$ satisfies the relation 
\begin{equation}
	2 \, G_{AB} \, R^B{}_C \, X^C = C_{ABC} \, R^B{}_D X^D \, R^C{}_E X^E \spc. 
	\label{cond2R} 
\end{equation}
Then, the additional first-order flow equations following from the stationarity of $S_2$ are 
\begin{subequations}
	\label{BPSflow5drot} 
	\begin{align}
		\tilde{\chi}^A &= 0 \;, \label{rot1}\\
		\partial_{\tau} w_4 &= 0 \;, \label{rot2}\\
		s N \partial_{\tau} w_5 &= p^0 w_5 + w_4 \;. 
	\end{align}
\end{subequations}

The coefficient $s$ is related to the sign of $W_5$, as follows. 
On the solution, we have {from} \eqref{rot1} that $\chi^A = - s f R^A{}_B \, X^B$, and both sides of this equation are a function of $\tau$. 
Differentiating $\chi^A$ with respect to $\tau$ and using the chain rule as well as \eqref{flowphi}, we obtain 
\begin{equation}
	\begin{split}
		\pd_\tau\chi^A &= -s \, R^A{}_B \left(-f^2\pd_\tau f^{-1} X^B + f\pd_\tau\phi^i\pd_i X^B\right) \\
		&= s\sgn(W_5) f^2 \, R^A{}_B \left(\frac{2}{3} X^B X^C + g^{ij}\pd_i X^B\pd_j X^C \right) Q_C \spc. 
	\end{split}
\end{equation}
Invoking the identity \eqref{ginvdXdX} we have 
\begin{equation}
	\pd_\tau\chi^A = s\sgn(W_5)f^2 G^{AB} q_B\spc, 
\end{equation}
and comparison with \eqref{flowchi} gives 
\begin{equation}
	s = \sgn(W_5) \spc. 
	\label{s} 
\end{equation}

Summarizing, we find that there are two classes of first-order flow equations specified by the sign of $W_5$. 
They are given by \eqref{BPSflow5d} and \eqref{BPSflow5drot}. 
Observe that \eqref{rot1} is the solution to \eqref{flowchi}, and that on a solution to \eqref{BPSflow5d}, the black hole potential \eqref{effpot} can also be written as 
\begin{equation}
	2 f^2 \left( \frac23 |W_5|^2 + g^{ij} \,\partial_i |W_5| \, \partial_j |W_5| \right) = 3 f^2 (\partial_{\tau} f^{-1})^2 + 2 g_{ij} \partial_{\tau} \phi^i \partial_{\tau} \phi^j \;. 
	\label{hamilton} 
\end{equation}
It can be checked that the five-dimensional Einstein-, Maxwell- and scalar field equations of motion derived from \eqref{action5} are satisfied for these two classes of flow equations.

The $X^A$ are taken to lie inside the K\"ahler cone (i.e. 
$X^A >0$). 
The flow equations \eqref{BPSflow5d} are then solved by \cite{Sabra:1997yd,Chamseddine:1998yv,Gauntlett:2004qy} (recall that $2V =1$) 
\begin{subequations}
	\label{solfiveext} 
	\begin{eqnarray}
		\label{solattr5} f^{-1} &=& \frac{2 }{3} \, H_A \, X^A \;, \\
		\chi^A &=& - s f \, R^A{}_B \, X^B \;, \label{solchi} \\
		f^{-1} \, X_A &=& \frac{1}{3} \, H_A \;, \label{XAH} 
	\end{eqnarray}
\end{subequations}
where $H_A$ denotes a harmonic function in three space dimensions, 
\begin{equation}
	H_A = h_A + |Q_A| \tau \;\;\;, \;\;\; h_A > 0 \;. 
\end{equation}
The $H_A$ are taken to be positive since this ensures that $f^{-1} > 0$ along the flow. 

The remaining flow equations of \eqref{BPSflow5drot} are solved by 
\begin{eqnarray}
	w_5 = H_0 = h_0 + q_0 \, \tau \;\;\;,\;\;\; w_4 = h^0 q_0 - h_0 p^0 \label{w5s1} 
\end{eqnarray}
for $s=1$, and by 
\begin{eqnarray}
	w_5 = c \, N^{-1} - \frac{w_4}{p^0} \;\;\;,\;\;\; {c = {\rm const}} \;\;\;,\;\;\; w_4 = {\rm const} \label{w5sm1} 
\end{eqnarray}
for $s=-1$. 
The $s=1$ solution is standard \cite{Gauntlett:2004qy}, and it describes a rotating supersymmetric solution in five dimensions provided that $Q_A = q_A$ (i.e.~$R^A{}_B = \delta^A{}_B$). 
The $s=-1$ solution, on the other hand, is non-standard. 
This solution is non-supersymmetric, since one of the conditions for supersymmetry derived in \cite{Gauntlett:2004qy}, namely that the self-dual part of $d w$ vanishes, is violated. 
In the absence of rotation, 
the five-dimensional solutions \eqref{solfiveext} are supersymmetric provided that $Q_A = q_A$ (i.e.~$R^A{}_B = \delta^A{}_B$ again).

Finally, let us comment on the boundary term in \eqref{stationarySBPS} which, on the solution \eqref{solfiveext}, equals $\frac23 f |W_5|$ evaluated at spatial infinity. 
This value is independent of both $p^0$ and the five-dimensional rotation parameter $w_5$. 
Therefore, it does not equal the ADM mass of the associated four-dimensional black hole which, in general, depends on both $p^0$ and $w_5$ (cf. 
\eqref{UfN}). 
The ADM mass of the $s=1$ solution \eqref{w5s1} is lower than the one of the $s=-1$ solution \eqref{w5sm1}.

\section{Flow equations in four dimensions \label{sec:flow4dw4}}

In the following, we relate the five-dimensional flow equations \eqref{BPSflow5d} and \eqref{BPSflow5drot} to four-dimensional ones by using the dictionary given in section \ref{secdic}. 
We set $w_4 =0$ in order to obtain static solutions in four dimensions. 
The four-dimensional flow equations then take the form 
\begin{eqnarray}
	\partial_{\tau} U &=& {\rm e}^U \, W_4 \;, \nonumber\\
	\partial_{\tau} z^A &=& 2 {\rm e}^U g^{A {\bar B}} \partial_{\bar B} W_4 \;, \label{4dflow} 
\end{eqnarray}
with a suitably identified $W_4$ as in \cite{Ceresole:2007wx,Andrianopoli:2007gt}. 
Observe that for a supersymmetric flow in four dimensions \cite{Ferrara:1997tw,Moore:1998pn,Denef:2000nb}, 
\begin{equation}
	- W_4 = \abs{Z_4} \;, \label{wz} 
\end{equation}
with $Z_4$ given in \eqref{central4}.

\subsection{Black holes with $w_5 =0$ \label{floweqst}}

In the absence of rotation in five dimensions we infer from \eqref{phi4} and \eqref{rel2u2phi} that the four-dimensional quantity ${\rm e}^{U}$ is expressed as 
\begin{eqnarray}
	{\rm e}^{-4 U} = {\rm e}^{4 \phi} \, f^{-4} = N \, f^{-3} \label{relUnf} 
\end{eqnarray}
in terms of five-dimensional quantities. 
Differentiating with respect to $\tau = 1/r$ and using \eqref{N} gives 
\begin{eqnarray}
	\partial_{\tau} {\rm e}^{-U} = \frac14 \left( {\rm e}^{3 U} \, f^{-3} \, p^0 + 3 \, {\rm e}^{\phi} \, \partial_{\tau} f^{-1} \right) \;. 
\end{eqnarray}
Using \eqref{phiK} we obtain 
\begin{eqnarray}
	\partial_{\tau} {\rm e}^{-U} = \frac18 \left( {\rm e}^{- K/2} \, p^0 + 12 \, {\rm e}^{K/2} \,{\rm e}^{- 2\phi} \, \partial_{\tau} f^{-1} \right) \;. 
	\label{flowU2} 
\end{eqnarray}
Inserting the five-dimensional flow equation for $f^{-1}$ given in \eqref{flowfi} into the above expression yields 
\begin{equation}
	\pd_\tau U = \ee^U \,W_4\spc, 
\end{equation}
with 
\begin{equation}
	W_4 = - \frac{1}{8} \, \ee^{K/2} \, \left( \ee^{-K} \, p^0 + 4 \, \left|Q_A (z^A - {\bar z}^A) \right| \right)\spc. 
	\label{W4} 
\end{equation}
Here we used that $C^A =0$ according to \eqref{casol}, so that $z^A = \im {\hat X}^A$.

Similarly, it can be checked that the flow equation for the scalar fields $z^A$ is given by \eqref{4dflow} with $W_4$ given by \eqref{W4}. 
This is done in appendix \ref{flowzfour}.

Setting $T^A = - \im z^A$ and taking the real part of $T^A$ to lie inside the K\"ahler cone, i.e. 
$T^A + {\bar T}^A >0$, yields (with $p^0 > 0$) 
\begin{equation}
	- W_4 = \frac{1}{8} \, \ee^{K/2} \, \left( \ee^{-K} \, p^0 + 4 \, \left| Q_A \, (T^A + {\bar T}^A) \right| \right) > 0 \spc, \label{W4t} 
\end{equation}
which is non-vanishing along the flow. 
This we now compare with the absolute value of the central charge $Z_4$ given in \eqref{z4}, 
\begin{equation}
	\abs{ Z_4} = \frac{1}{8} \, \ee^{K/2} \, \left| \ee^{-K} \, p^0 + 4 \, q_A \, (T^A + {\bar T}^A) \right| \spc. 
\end{equation}
Both expressions only agree provided that $Q_A (T^A + {\bar T}^A) > 0$ and $Q_A = q_A$, in which case the flow is supersymmetric in four dimensions, since it is derived from $\abs{Z_4}$. 
Otherwise the flow is non-supersymmetric. 
First-order flow equations based on \eqref{W4t} were first obtained in \cite{Ceresole:2007wx} in the context of the STU-model (which corresponds to $C_{123} =1$) by using a different approach. 

The solution to the first-order flow equations \eqref{4dflow} based on \eqref{W4t} reads 
\begin{eqnarray}
	\ee^{-2U} &=& \frac23 N^{1/2} \, H_A \; f^{-1/2} \, X^A \;, \nonumber\\
	z^A &=& \im {\hat X}^A = \im N^{-1/2} \; f^{-1/2} \, X^A \;, \label{solw5zero} 
\end{eqnarray}
where $f^{-1/2} \, X^A$ is the solution to \eqref{XAH}, and where we used \eqref{relUnf}. 
This solution was first derived in \cite{Kallosh:2006ib} for the STU-model by solving the four-dimensional equations of motion. 
The horizon is at $\tau = \infty$, and the scalar fields $z^A$ get attracted to constant values there. 
Computing $\ee^{-2U}$ at the horizon yields the entropy of the extremal black hole 
\begin{equation}
	{\cal S} = \pi \, \abs{W_4}^2_{\rm hor} \label{entrow4} 
\end{equation}
in accordance with \cite{Ceresole:2007wx,Andrianopoli:2007gt}. 
An equivalent expression for the entropy can be found in \cite{Sahoo:2006rp,Cardoso:2006xz}.

We may apply duality transformations to obtain first-order flow equations for extremal non-supersymmetric solutions in four dimensions carrying other types of charges. 
This was also discussed in \cite{Kallosh:2006ib}. 
We consider the prepotential (we refer to appendix \ref{sec:VSG} and to \cite{Behrndt:1996jn} for some of the conventions used) 
\begin{equation}
	F(Y) = - \frac16 \, \frac{C_{ABC} \, Y^A Y^B Y^C}{Y^0} \;, 
\end{equation}
and apply, for instance, the non-perturbative duality transformation $(Y^I, F_I) = (-{\tilde F}_I, {\tilde Y}^I)$. 
This gives rise to extremal black hole solutions of the type recently discussed in \cite{Tripathy:2005qp,Kallosh:2006bt,Nampuri:2007gv}, as follows. 
The $z^A= Y^A/Y^0$ can then be expressed as $z^A = {\tilde F}_A/{\tilde F}_0$ and the charges $(p^0, Q_A)$ become equal to $(- {\tilde q}_0, {\tilde P}^A)$ (and similarly for the respective harmonic functions). 
Observe that since $Q_A = q_B \, R^B{}_A $, the charge ${\tilde P}^A$ does not equal ${\tilde p}^A = q_A$ in general. 
Since the combination $\im \left( {\bar Y}^I \, F_I - Y^I \, {\bar F}_I \right)$ is invariant under symplectic transformations, and because it is equal to $\abs{Y^0}^2 \, \ee^{-K}$, it follows that 
\begin{equation}
	\abs{{\tilde Y}^0}^2 \, \ee^{- {\tilde K}} = \abs{Y^0}^2 \, \ee^{- K} \;. 
	\label{kktilde} 
\end{equation}
Hence we can write 
\begin{equation}
	\frac{\ee^{{\tilde K}/2}}{\abs{{\tilde Y}^0}} = \frac{\ee^{K/2}}{\abs{Y^0}} = \frac{\ee^{K/2}}{\abs{{\tilde F}_0}} \;. 
	\label{kktilderel} 
\end{equation}
Also, using $z^A = \im {\hat X}^A$ as well as \eqref{prep4} and \eqref{emK}, we have $\abs{F_0} = \frac18 \abs{Y^0} \, \ee^{-K}$. 
Then, we find that \eqref{W4t} can be expressed in terms of the transformed quantities as 
\begin{eqnarray}
	- {W}_4 &=& \frac{1}{8} \, \frac{\ee^{{K}/2}}{\abs{{\tilde F}_0}} \, \left( - \ee^{- {K}} \, \abs{{\tilde F}_0} \, {\tilde q}_0 + 8 \, \abs{{\tilde P}^A \, {\tilde F}_A} \right) \nonumber\\
	&=& \ee^{{\tilde K}/2} \left( - {\tilde q}_0 + \, \left|{\tilde P}^A \, \frac{{\tilde F}_A}{{\tilde Y}^0} \right| \right) \spc. 
	\label{w4q0} 
\end{eqnarray}
Observe that $z^A = - {\bar z}^A$ implies that also ${\tilde z}^A = - {\bar {\tilde z}}^A$. 
Thus, extremal four-dimensional black hole solutions with charges $({\tilde q}_0, {\tilde P}^A)$ and with scalar fields satisfying ${\tilde z}^A = - {\bar {\tilde z}}^A$ can be obtained by solving the first-order flow equations \eqref{4dflow} based on \eqref{w4q0}. 
Observe that $- {\tilde q}_0 = p^0 > 0$ and ${\tilde F}_A/{\tilde Y}^0 = - Y^A/F_0 > 0$. 
Hence, only when ${\tilde P}^A = {\tilde p}^A$ and ${\tilde P}^A {\tilde F}_A/{\tilde Y}^0 >0$ is $-W_4 = \abs{Z_4}$, as can be seen from \eqref{central4}, and the resulting flow is supersymmetric. 
Otherwise, the flow is non-supersymmetric.

\subsection{Black holes with $w_5 \neq 0$ \label{floweqrot}}

Now we derive the four-dimensional flow equations associated with rotating black holes with $w_5 \neq 0$ in five dimensions. 
Since in the presence of rotation, the matrix $R^A{}_B$ has to satisfy the additional constraint \eqref{cond2R}, we take $R^A{}_B = \delta^A{}_B $ for simplicity. 
Other $R^A{}_B$ satisfying \eqref{cond2R} may also exist. 
This will be analyzed elsewhere.

{From} \eqref{phi4} and \eqref{rel2u2phi} we obtain 
\begin{equation}
	\ee^{2 U} = \ee^{2 \phi} \, f \, N^{-1} = \ee^{-2 \phi} \, f^2 \, \Delta \;, \label{uw5} 
\end{equation}
where 
\begin{equation}
	\Delta = 1 + \ee^{4 \phi} \left(f \, w_5 \right)^2 \;. 
	\label{deltw5} 
\end{equation}
Using \eqref{uw5} we have 
\begin{equation}
	\ee^{-4U} = N \,f^{-3} - (N \, w_5)^2 \label{UfN} 
\end{equation}
as well as 
\begin{eqnarray}
	\ee^{3U} \, f^{-3} &=& \ee^{- 3 \phi} \, \Delta^{3/2} \;, \nonumber\\
	\ee^{3 U} \, N \, f^{-2} &=& \ee^{\phi} \, \Delta^{1/2} \;. 
	\label{relUfDel} 
\end{eqnarray}
Combining \eqref{casol} with \eqref{solchi} results in 
\begin{equation}
	C^A = - s \, w_5 \, f \, \ee^{2 \phi} \, {\hat X}^A \;, 
\end{equation}
and hence 
\begin{equation}
	z^A = \alpha \, {\hat X}^A \;\;\;,\;\;\; \alpha = -s \, w_5 \, f \, \ee^{2 \phi} + \im \;. 
	\label{valueza} 
\end{equation}
Comparing \eqref{deltw5} with \eqref{valueza} we obtain 
\begin{equation}
	\Delta = |\alpha|^2 \;\;\;,\;\;\; 4 - 3 \Delta = 1 - 3 \left( {\rm Re} \alpha \right)^2 \;. 
	\label{deltaal} 
\end{equation}

We begin by considering the $s=1$ solution given by \eqref{w5s1}, with $w_4 = 0$, which implies 
\begin{equation}
	w_5 = H_0 = q_0 N/p^0 \;. 
	\label{H0N} 
\end{equation}
{From} \eqref{relUfDel} and \eqref{deltw5} we have 
\begin{eqnarray}
	\ee^{3 U} \, N \, w_5^2 = \ee^{-3 \phi} \, \Delta^{1/2} \, (\Delta -1) \;. 
	\label{hzerofdel} 
\end{eqnarray}
Using \eqref{UfN}, \eqref{H0N} and the flow equation \eqref{flowfi}, we obtain 
\begin{eqnarray}
	\partial_{\tau} \, \ee^{-U} = \frac14 \ee^{3U} \left( p^0 f^{-3} + 2 N f^{-2} W_5 - 4 p^0 \, N \, H_0^2 \right) \;. 
	\label{dtu} 
\end{eqnarray}
Inserting \eqref{relUfDel} and \eqref{hzerofdel} into \eqref{dtu} results in 
\begin{equation}
	\partial_{\tau} \, \ee^{-U} = \frac14 \, \ee^{ 3 \phi} \, \Delta^{1/2} \left( p^0 \, \ee^{-6 \phi} \, (4 - 3 \Delta) + 2 q_A \, {\hat X}^A \right) \;. 
	\label{dtuw} 
\end{equation}
The right hand side of \eqref{dtuw} is identified with the ``superpotential'' 
\begin{equation}
	- W_4 = \frac14 \, \ee^{3 \phi} \, \Delta^{1/2} \left[ p^0 \, \ee^{-6 \phi} \, \left( 1 - 3 \left( {\rm Re} \alpha \right)^2 \right) + 2 q_A \, {\hat X}^A \right] \;. 
	\label{w4s1} 
\end{equation}
This we compare with $Z_4$ given in \eqref{z4}, which for the case at hand reads 
\begin{equation}
	Z_4 = \frac12 \, \ee^{K/2} \, \alpha \, \left[ - p^0 \, \ee^{- 6 \phi} \left( 1 - 3 \left( {\rm Re} \alpha \right)^2 \right) - 2 q_A \, {\hat X}^A \right] \;. 
\end{equation}
Using \eqref{phiK} we find that \eqref{wz} holds.
Thus, the $s=1$ flow is a supersymmetric flow in four dimensions. 
The solution to the first-order flow equations \eqref{4dflow} is well known and reads 
\begin{eqnarray}
	\ee^{-4U} &=& \frac49 \,N \, \left( H_A \; f^{-1/2} \, X^A \right)^2 - \left(N \, H_0 \right)^2\;, \nonumber\\
	z^A &=& \alpha \,{\hat X}^A = \frac32 \, \left(\frac{- N \, H_0 + \im \ee^{-2U} }{N \, H_B \; f^{-1/2} \, X^B } \right) \; f^{-1/2} \, X^A \;, \label{sol4s1} 
\end{eqnarray}
where $f^{-1/2} \, X^A$ is the solution to \eqref{XAH}. 

Next, we consider the $s=-1$ solution given in \eqref{w5sm1}, with $w_4 =0$. 
{From} \eqref{UfN} we obtain 
\begin{equation}
	\ee^{-4U} = N \, f^{-3} - c^2 \;, 
\end{equation}
which we demand to be positive at spatial infinity (i.e. 
at $\tau =0$) to ensure that $\ee^{-4U}$ remains non-vanishing along the flow. 
Using the flow equation \eqref{flowfi} as well as \eqref{relUfDel} we get 
\begin{eqnarray}
	\partial_{\tau} \, \ee^{-U} &=& \frac14 \ee^{3U} \left( p^0 f^{-3} - 2 N f^{-2} \, W_5 \right) \nonumber\\
	&=& \frac14 \ee^{3 \phi} \, \Delta^{1/2} \left( \ee^{- 6 \phi} \, \Delta \, p^0 - 2 q_A \, {\hat X}^A \right) \;. \label{w4smone} 
\end{eqnarray}
With the help of \eqref{valueza} and \eqref{deltaal} this can be rewritten as 
\begin{eqnarray}
	\partial_{\tau} \, \ee^{-U} = - W_4 
\end{eqnarray}
with 
\begin{eqnarray}
	- W_4 = \ee^{K/2} \, \left| \frac{p^0}{6} \, C_{ABC} z^A \, z^B \, {\bar z}^C - q_A \, z^A \right| \;. 
	\label{w4zzbz} 
\end{eqnarray}
Similarly, one can verify that the flow equation for the scalar fields $z^A$ is given by \eqref{4dflow} with $W_4$ as in \eqref{w4zzbz}. 
This is done in appendix \ref{flowzfour}. 
Thus, we find that the $s=-1$ solution is described by a first-order flow equation based on \eqref{w4zzbz}. 
Inspection of \eqref{w4smone} shows that $W_4$ is non-vanishing along the flow. 
An example of a flow of this type was constructed recently in \cite{Ceresole:2007wx}, where also the stability of the solution is discussed. 

The solution to the first-order flow equation \eqref{4dflow} based on \eqref{w4zzbz} reads 
\begin{eqnarray}
	\ee^{-4U} &=& \frac49 \,N \, \left( H_A \; f^{-1/2} \, X^A \right)^2 - c^2 \;, \nonumber\\
	z^A &=& \alpha \,{\hat X}^A = \frac32 \, \left(\frac{c + \im \ee^{-2U} }{N \, H_B \; f^{-1/2} \, X^B } \right) \, f^{-1/2} \, X^A \;, \label{solw5non} 
\end{eqnarray}
where $f^{-1/2} \, X^A$ is the solution to \eqref{XAH}. 
When setting $c=0$, the solution \eqref{solw5non} reduces to the static case \eqref{solw5zero}. 
The solution \eqref{solw5non} is such that the axions $C^A$ vanish at the horizon (i.e.~at $\tau = \infty$), so that $z^A_{\rm hor} = \im {\hat X}^A_{\rm hor}$, and the entropy is given by \eqref{entrow4}. 
Away from the horizon the $C^A$ are non-vanishing and therefore the axions are subject to a non-trivial flow.

Finally, we may apply the non-perturbative duality transformation discussed below \eqref{entrow4} to \eqref{w4zzbz}. 
Using \eqref{prep4} we have $F_0 = \frac18 \, \alpha^3 \, Y^0 \, \ee^{-K}$, and hence 
\begin{equation}
	- W_4 = \ee^{{\tilde K}/2} \, \left| \frac{\bar \alpha}{\alpha} \, {\tilde q}_0 - {\tilde p}^A \, \frac{{\tilde F}_A}{{\tilde Y}^0} \right| \;. 
	\label{w4alp} 
\end{equation}
The phase ${\bar \alpha}/\alpha$ can be expressed in terms of the transformed quantities as follows. 
For the non-perturbative duality transformation under consideration, ${\tilde F}({\tilde Y}) = F(Y)$ (see for instance \cite{deWit:1996ix}). 
Since $({\bar \alpha}/\alpha)^3 = {\bar F} ({\bar Y})/F(Y)$, it follows 
\begin{equation}
	\left(\frac{\bar \alpha}{\alpha} \right)^3 = \frac{{\bar {\tilde F}}({\bar {\tilde Y}})}{{\tilde F}({\tilde Y})} \;. 
\end{equation}

\section{Multiple $W_5$ for a given black hole potential \label{sec:W5V}}

In this section we extend our discussion of section \ref{sec:flow5dw5} on the black hole potential and its description in terms of $W_5$.

One of the key features that allow to identify classes of four-dimensional stable extremal black holes described by first-order differential equations in \cite{Ceresole:2007wx} is the degenerate description of the four-dimensional effective potential in terms of a ``superpotential'' $W_4$. 
The solutions are supersymmetric only when the latter coincides with the four-dimensional central charge $Z_4$.

A similar analysis can be carried out in five dimensions. 
As discussed in section \ref{sec:flow5dw5}, whenever the five-dimensional effective potential is expressed in terms of a ``superpotential'' $W_5$ as in \eqref{effpot}, first-order flow equations for the various fields describing the extremal black hole can be obtained. 
The associated solutions may be supersymmetric in five dimensions when $W_5$ equals the five-dimensional central charge $Z_5 = q_A \, X^A$.

In section \ref{sec:flow5dw5} we focused on $W_5$'s which are obtained by studying the invariance group of the inverse matrix $G^ {AB}$ (see \eqref{condition}). 
This is similar to the discussion in \cite{Ceresole:2007wx} of the invariances of the complex matrix ${\cal M}$ appearing in the four-dimensional black hole potential, only that it is simpler in five dimensions because $G^{AB}$ is real. 
It may be useful to note that although the matrices $R^A{}_B$ are part of the invariance group of the norm defined by the inverse metric $G^{AB}$, they can also be interpreted as transformations on the moduli space of very special geometry by using relations such as (\ref{gij}).

Rather than attempting to characterize the general form of such matrices $R^A{}_B$, let us in the following discuss a few classes of very special geometries for which it is possible to find non-trivial solutions to (\ref {condition}). 
Generically, $G^{AB}$ (with $A,B = 1, \dots, n$) has non-zero entries in all of its matrix elements and possesses $n$ different eigenvalues. 
Then, the only allowed matrix $R^A{}_B$ is the identity matrix. 
Non-trivial solutions can be found when $G^{AB}$ has $m$ identical eigenvalues, in which case $R^A{}_B$ is an orthogonal matrix in $O(m)$. 
A further specialization arises when $G^{AB}$ becomes (block) diagonal. 
Consider, for instance, the class of scalar manifolds 
\begin{equation}
	{\cal M} = \frac{SO(n-1,1)}{SO(n-1)} \times SO(1,1) \;, \label{class} 
\end{equation}
which is associated to the Jordan algebra $J = \Sigma_{n-1} \times {\mathbb R}$, where $\Sigma_{n-1}$ is the Jordan algebra of degree two corresponding to a quadratic form of signature $(+ - \ldots -)$. 
The inverse $G^{AB}$ therefore factorizes into a generic $(n-1) \times (n-1)$-block and a single entry for the extra $SO(1,1)$ factor \cite{Gunaydin:1983bi}. 
This means that the associated ``superpotential'' can be chosen as 
\begin{equation}
	W_5 = \pm q_1 X^1 + q_a X^a \;\;\;,\;\;\; a= 2, \dots, n \;, \label{supocase1} 
\end{equation}
where $X^1$ is the ambient vector space coordinate associated with the $SO(1,1)$ factor. 
The exceptional case $n=2$ with moduli $a$ and $b$ has a metric which further degenerates to a diagonal form and therefore admits arbitrary sign changes in front of any of the three charges allowed by the model in the ``superpotential''. 
Let us discuss this in more detail. 
The scalar manifold is simply $SO(1,1) \times SO(1,1)$ and it can be obtained as a hypersurface in an ambient vector space, parameterized by the $X^A$ coordinates. 
The matrix $G_{AB}(X^C)$ is the metric of this ambient space and has non-trivial entries in all of its elements as functions of $X^C$. 
After the restriction to the hypersurface characterized by $2 V = 1$, this metric reduces to the diagonal form 
\begin{equation}
	G_{AB} = \left( 
	\begin{array}{ccc}
		\frac{1}{a^2} & 0 &0\\
		0 & \frac{1}{b^2} & 0\\
		0 & 0 & \frac{a^2 b^2}{2} 
	\end{array}
	\right). 
	\label{metrica} 
\end{equation}
{From} \eqref{metrica} it is clear that any diagonal matrix $R^A{}_B$ with entries $\pm 1$ solves \eqref{condition} and therefore leaves the black hole potential invariant. 
The various ``superpotentials'' are then defined by $W_5 = q_A R^A{}_B X^B $, and an example thereof is \eqref{supocase1}. 
Only when the matrix $R^A{}_B$ is the identity is it possible to obtain supersymmetric solutions.

In order to obtain further insights into the degenerate description of the black hole potential in terms of ``superpotentials'', additional guidance beyond the discussion of invariances of the inverse metric $G^{AB}$ is needed. 
It is known that group theoretical tools, such as the analysis of orbits of U-duality groups in $N \geq 2$ supergravity theories in four and five dimensions \cite{Ferrara:1997uz, Ferrara:2006xx,Bellucci:2006xz}, can be used to classify both BPS and non-BPS states. 
U-duality can also be used to shed light on the degenerate description of the black hole potential in terms of ``superpotentials'' $W$. 
In the recent work \cite{Andrianopoli:2007gt} it was shown that in four-dimensional supergravity theories with $N>2$, some of the $W$'s giving rise to the black hole potential can be written in terms of linear combinations of U-duality invariants. 
Both the supersymmetric and non-supersymmetric critical points of the black hole potential are then related to its different rewritings in terms of these invariants. 
It remains to be seen whether these results can be extended to the case of general $N=2$ supergravity theories in four and five dimensions.

\section{Summary}

In this note we constructed single-center static extremal non-supersymmetric black hole solutions in four-dimensional $N=2$ supergravity theories based on general cubic prepotentials. 
These solutions have full radial dependence and interpolate between Minkowski spacetime at spatial infinity and the near-horizon geometry. 
The solutions constructed here include the ones found in \cite{Kallosh:2006ib} in the context of the STU-model. 
The construction given in this note, however, applies to any cubic prepotential and is based on first-order flow equations, rather than on the equations of motion as was the case in \cite{Kallosh:2006ib}.

First-order flow equations exist when the effective black hole potential can be expressed in terms of a ``superpotential'' $W$ \cite{Ceresole:2007wx}. 
The rewriting of the black hole potential in terms of $W$ is, however, not unique. 
A given black hole potential may thus give rise to different first-order flow equations and it is only in special cases that these match the supersymmetry conditions. 
Hence the corresponding solutions describe black holes that may or may not be supersymmetric, depending on the choice of $W$. 
Following this approach, examples of first-order flow equations for extremal non-supersymmetric black hole solutions in four dimensions were discussed in \cite{Ceresole:2007wx,Andrianopoli:2007gt}. 
In this note, on the other hand, we gave a systematic procedure for constructing extremal non-supersymmetric black hole solutions in four dimensions by using the connection between five- and four-dimensional extremal black holes. 
This 5d/4d-connection relates static solutions in four dimensions to rotating solutions in five dimensions and has been used in the past both in the context of supersymmetric black holes and in the context of the entropy function approach for determining the near-horizon solution of extremal black holes (see for instance \cite{Behrndt:2005he} and \cite{Cardoso:2007rg}, respectively). 
Here we used the connection with five dimensions to construct four-dimensional ``superpotentials'' $W_4$ in terms of five-dimensional data. 
In fact, the 5d/4d-connection allows for a rather simple construction of $W_4$, as follows. 
The first step consists in the rewriting of the five-dimensional black hole potential \eqref{effpot} in terms of a five-dimensional ``superpotential'' $W_5$, which need not be identical to the five-dimensional central charge $Z_5$ (this possibility follows again by noting that the rewriting of the black hole potential in terms of a single scalar function is not unique, as detailed in section \ref{sec:W5V}). 
The $W_5$ we employed is expressed in terms of the electric charges $q_A$ and a constant real matrix $R^A{}_B$ associated with the invariance group of $G^{AB}$, see \eqref{w5}. 
This ``superpotential'' then leads to first-order flow equations for extremal black holes in a Taub-NUT geometry in five dimensions. 
In the static case, the flow equations are given by \eqref{BPSflow5d}, and they are solved by \eqref{solfiveext}. 
In the rotating case, the flow equations are given by \eqref{BPSflow5d} and \eqref{BPSflow5drot}, and are supplemented by the condition \eqref{cond2R}. 
In the rotating case, the stationary part $w_5$ of the five-dimensional line element is determined by the sign $s$ of $W_5$, which in turn is determined by the sign of the ``rotated'' charges $Q_A = q_B \, R^B{}_A$ appearing in $W_5$. 
Depending on the sign $s$, the solution for $w_5$ takes a different form. 
When $s = 1$, $w_5$ takes the standard form \eqref{w5s1} in terms of harmonic functions, also known from the study of rotating supersymmetric solutions \cite{Gauntlett:2004qy}. 
When $s = -1$, however, $w_5$ takes the new non-standard form \eqref{w5sm1}.

Reducing the static five-dimensional solutions down to four dimensions yields the static extremal black hole solutions \eqref{solw5zero}, which are valid for any matrix $R^A{}_B$ satisfying \eqref{condition}.

On the other hand, when reducing the rotating five-dimensional solutions down to four dimensions, one needs to take into account the additional constraint \eqref{cond2R}. 
In this note we took $R^A{}_B = \delta^A{}_B$ for simplicity in the rotating case, but other solutions to \eqref{cond2R} may exist and will be discussed elsewhere. 
The resulting four-dimensional solutions are labelled by the sign $s$ which, for the case at hand, is related to the sign of the electric charges $q_A$. 
When $s=1$, the resulting solution is supersymmetric and takes the well known form \eqref{sol4s1}, whereas when $s=-1$ the solution is non-supersymmetric and is given by \eqref{solw5non}.

The extremal solutions presented in this note are solutions to any $N=2$ supergravity theory based on a cubic prepotential. 
They are for the most part new and include, as special subcases, the $N=2$ solutions discussed in \cite{Kallosh:2006ib,Ceresole:2007wx,Andrianopoli:2007gt}. 
Other extremal solutions may then be generated by applying electric-magnetic duality transformations to them (see for instance \eqref{w4q0} and \eqref{w4alp}). 
We note, however, that the construction of extremal black hole solutions described in this note does not 
capture the entire class of such solutions.  For instance, it does not allow for the construction
of an extremal non-supersymmetric solution with non-vanishing charges $(q_0, p^0)$.

Finally, it is worth pointing out that only a subset of the extremal non-supersymmetric four-dimensional black holes which we constructed as solutions to first-order flow equations are connected to supersymmetric solutions in five dimensions. 
Hence only in certain cases can one argue that the first-order flow equations are due to hidden supersymmetry as in \cite{Nilsson:1984bj,Duff:1997qz}. 
The solutions found in this paper show that the presence of first-order flow equations is a feature that is more general and not necessarily tied to the presence of supersymmetry. 
It would be interesting to study this further and, in particular, to see if these first-order equations can be recovered in the context of fake supergravity \cite{Freedman:2003ax,Celi:2004st}.

\subsection*{Acknowledgements}

This work is supported by the European Union contract MRTN-CT-2004-005104 ``Constituents, Fundamental Forces and Symmetries of the Universe''.

\appendix

\section{Review of very special geometry\label{sec:VSG}}

The five-dimensional $N=2$ supergravity action is based on the cubic polynomial \cite{Gunaydin:1983bi} 
\begin{equation}
	V = \frac{1}{6} C_{ABC} X^A X^B X^C \;, \label{constrV} 
\end{equation}
where the $X^A (\phi)$ are real scalar fields satisfying the constraint $V = {\rm constant}$. 
In our conventions $2 V = 1$ \cite{Cardoso:2007rg}.

{From} the definitions 
\begin{gather}
	X_A = \frac{1}{6} C_{ABC} X^B X^C \spc, \label{Xdual} \\
	G_{AB} = \frac{1}{V}\left(-\frac{1}{2}C_{ABC} X^C + \frac{9}{2}\frac{X_A X_B}{V}\right) \label{GAB} \;, 
\end{gather}
it follows that 
\begin{equation}
	X^A X_A = V 
\end{equation}
and 
\begin{equation}
	X_A = \frac{2V}{3} G_{AB} X^B \spc, \qquad X^A = \frac{3}{2V} G^{AB} X_B \spc, \label{lowerraiseA} 
\end{equation}
where $G_{AB} \, G^{BC} = \delta_A^C$.

As $V$ is a constant, i.e.\ $\pd_i V = 0$ and $\pd_i V = \frac{3}{2}(\pd_i X_A) X^A$, we also have 
\begin{equation}
	(\pd_i X_A) X^A = 0 \spc \label{dXAXA} 
\end{equation}
and consequently, by the definitions \eqref{Xdual} and \eqref{GAB}, 
\begin{equation}
	G_{AB}\pd_i X^B = -\frac{3}{2V}\pd_i X_A \spc. 
	\label{GABdXB} 
\end{equation}

The metric on the scalar manifold is defined by 
\begin{equation}
	g_{ij} = G_{AB} \pd_i X^A \pd_j X^B \spc. 
	\label{gij} 
\end{equation}
The index structure dictates that 
\begin{equation}
	g^{ij} \pd_i X^A \pd_j X^B = a(G^{AB} - b X^A X^B) \label{ginvdXdXansatz} 
\end{equation}
with constant coefficients $a$ and $b$. 
Contraction with $X_B$ must vanish because of eq.~\eqref{dXAXA}. 
>From eq.~\eqref{lowerraiseA} we then have 
\begin{equation}
	0 = g^{ij} \pd_i X^A \pd_j X^B X_B = a(\ft{2}{3}V X^A - b X^A V) \spc, 
\end{equation}
which fixes $b = \frac{2}{3}$. 
To determine the coefficient $a$ we contract eq.~\eqref{ginvdXdXansatz} with $G_{AB}$, invoke the definition \eqref{gij} and observe that the number of physical scalars is one less than the number of vector fields, $n$, 
\begin{equation}
	n-1 = g_{ij}g^{ij} = G_{AB} g^{ij} \pd_i X^A \pd_j X^B = a(G_{AB} G^{AB} - \ft{2}{3} G_{AB} X^A X^B) = a(n-1) \spc. 
\end{equation}
This implies that $a=1$, so that finally 
\begin{equation}
	g^{ij} \pd_i X^A \pd_j X^B = G^{AB} - \frac{2}{3} X^A X^B \spc. 
	\label{ginvdXdX} 
\end{equation}

The four-dimensional $N=2$ supergravity action corresponding to \eqref{constrV} is based on the prepotential \cite{deWit:1984pk,deWit:1984px} 
\begin{equation}
	F(Y) = - \frac16 \, \frac{C_{ABC} \, Y^A Y^B Y^C}{Y^0} \;, \label{prep4} 
\end{equation}
where the $Y^I$ are complex scalar fields ($I = 0, A$). 
The four-dimensional physical scalar fields $z^A$ are 
\begin{equation}
	z^A = \frac{Y^A}{Y^0} \;. 
\end{equation}
The four-dimensional K\"ahler potential $K(z, {\bar z})$ derived from \eqref{prep4} is 
\begin{eqnarray}
	{\rm e}^{- K } = \frac{\im}{6} C_{ABC} \, (z^A - {\bar z}^A) \, (z^B - {\bar z}^B) \, (z^C - {\bar z}^C) \;. 
	\label{emK} 
\end{eqnarray}
The K\"ahler metric $g_{A {\bar B}} = \frac{\partial}{\partial z^A} \frac{\partial}{\partial {\bar z}^B} K$ derived from \eqref{emK} satisfies the relation 
\begin{equation}
	g_{A\bar{B}} = \frac12 \, \ee^{4\phi} \, G_{AB} \;. 
	\label{Gg} 
\end{equation}
The four-dimensional quantity $Z(Y)$ is given by 
\begin{equation}
	Z(Y) = p^I \, F_I (Y) - q_I \, Y^I \;, 
\end{equation}
where $F_I = \partial F(Y)/\partial Y^I$. 
The associated four-dimensional complex central charge $Z_4$ reads 
\begin{equation}
	Z_4 = \ee^{K/2} \, Z(Y)/Y^0 \;. 
	\label{central4} 
\end{equation}
For a prepotential of the form \eqref{prep4} we obtain 
\begin{equation}
	Z(Y) = Y^0 \left( \frac{p^0}{6} \, C_{ABC} z^A z^B z^C - \frac{p^A}{2} \, C_{ABC} z^B z^C - q_0 - q_A z^A \right) \;, 
\end{equation}
and hence, 
\begin{equation}
	Z_4 = {\rm e}^{K/2} \left( \frac{p^0}{6} \, C_{ABC} z^A z^B z^C - \frac{p^A}{2} \, C_{ABC} z^B z^C - q_0 - q_A z^A \right) \;. 
	\label{z4} 
\end{equation}

\section{Evaluating the action in five dimensions \label{bulkaction5}}

The square root of the determinant of the metric \eqref{metric}, \eqref{tnhyper} is 
\begin{equation}
	\sqrt{-G} = \frac{R r^2 N \sin\theta}{f} \;. 
\end{equation}
The inverse metric reads 
\begin{equation}
	G^{MN} = 
	\begin{pmatrix}
		-1/f^2 + f\left(\frac{N w_5^2}{R^2} + \frac{w_4^2}{r^2 N}\frac{\cos^2\theta}{\sin^2\theta}\right) & 0 & 0 & -\frac{f w_4}{r^2 N}\frac{\cos\theta}{\sin^2\theta} & f\left(-\frac{N w_5}{R^2} + \frac{p^0 w_4}{r^2 N}\frac{\cos^2\theta}{\sin^2\theta}\right) \\
		0 & f/N & 0 & 0 & 0 \\
		0 & 0 & f/(r^2 N) & 0 & 0 \\
		-\frac{f w_4}{r^2 N}\frac{\cos\theta}{\sin^2\theta} & 0 & 0 & \frac{f}{r^2 N}\frac{1}{\sin^2\theta} & -\frac{p^0 f}{r^2 N}\frac{\cos\theta}{\sin^2\theta} \\
		f\left(-\frac{N w_5}{R^2} + \frac{p^0 w_4}{r^2 N}\frac{\cos^2\theta}{\sin^2\theta}\right) & 0 & 0 & -\frac{p^0 f}{r^2 N}\frac{\cos\theta}{\sin^2\theta} & f\left(\frac{N}{R^2} + \frac{(p^0)^2}{r^2 N}\frac{\cos^2\theta}{\sin^2\theta}\right) 
	\end{pmatrix}
\end{equation}
and the Ricci scalar is 
\begin{equation}
	\label{Ricci} 
	\begin{split}
		\mathcal{R} = \frac{1}{2 r^4 f N^3}\biggl[ & f^5 N \left((p^0 w_5 + w_4)^2 + \frac{r^4 N^2}{R^2} w_5'^2 + r^2 \cot^2\theta\,w_4'^2\right) \\
		& - 5 r^4 N^2 f'^2 + 2 r^3 f N^2 (2 f' + r f'') \\
		& - f^2 \left( (p^0)^2 R^2 - r^4 N'^2 + 2 r^3 N (2 N' + r N'') \right) \biggr] , 
	\end{split}
	\raisetag{5ex} 
\end{equation}
where the prime denotes the derivative with respect to the radial coordinate $r$, i.e. 
$' = \pd/\pd r$. 
The last line in the expression \eqref{Ricci} above vanishes on account of the definiton \eqref{N}.

Inserting the ansatz 
\begin{eqnarray}
	A^A_5 = \chi^A (r) \, ( dt + w) + p^A \, \cos \theta \, d\varphi 
\end{eqnarray}
into the gauge kinetic term in \eqref{action5} yields 
\begin{equation}
	\begin{split}
		-\frac{1}{2}\sqrt{-G}\, G_{AB} F^A_{MN} F^{BMN} = &-\frac{R \sin\theta}{r^2 f^2 N} G_{AB} \Bigl[f^3 p^A p^B + f^3 (p^0 w_5 + w_4)(p^A \chi^B + \chi^A p^B) \\
		&+ f^3\Bigl((p^0 w_5 + w_4)^2 + \frac{r^4 N^2}{R^2} w_5'^2 + r^2 \cot^2\theta\, w_4'^2\Bigr)\chi^A \chi^B \\
		&- r^4 N \chi'^A \chi'^B \Bigr] \;. 
	\end{split}
\end{equation}
The Chern--Simons term in \eqref{action5} evaluates to 
\begin{eqnarray}
	- \frac{1}{6V} C_{ABC} \, F^A \wedge F^B \wedge A^C &=& \frac{1}{3V} \sin\theta\, C_{ABC} \left[p^A + (p^0 w_5 + w_4) \chi^A \right] \chi^B \chi^C w_5' \nonumber\\
	&& \hskip 1cm dt\wedge dr\wedge d\theta \wedge d\varphi \wedge d\psi \;. 
\end{eqnarray}

\section{Flow equations for the complex scalar fields $z^A$ \label{flowzfour}}

Here, we derive the flow equation for the complex scalars $z^A$. 
We set $2 V = 1$. 
We begin by first considering the case discussed in subsection \ref{floweqst}, so that $z^A = \im {\hat X}^A$. 
Using \eqref{relUnf} we obtain 
\begin{equation}
	\pd_\tau\Xh^A = \pd_\tau(f^{-2}\ee^{2U})X^A + \ee^{-2\phi} \, \pd_\tau\phi^i \, \pd_i X^A \spc, \label{dtauXhorig} 
\end{equation}
where $\pd_i$ stands for the derivatives with respect to the physical scalars $\phi^i$ in five dimensions. 
{From} \eqref{flowfi} and \eqref{flowU2} we get 
\begin{equation}
	\begin{split}
		\label{f2u} \pd_\tau(f^{-2}\ee^{2U}) &= \frac43 \ee^{2U} f^{-1}\abs{Q_A X^A} - 2f^{-2}\ee^{3U}\pd_\tau\ee^{-U} \\
		&= \frac43 f\abs{Q_A \Xh^A} - \frac{1}{4}f^{-2}\ee^{3U}\left(\ee^{-K/2} p^0 + 4\ee^{K/2}\abs{Q_A \Xh^A}\right) \spc. 
	\end{split}
\end{equation}
Using the flow equation for $\phi^i$ given in \eqref{flowphi} we obtain 
\begin{equation}
	\partial_{\tau} \phi^i \, \partial_i X^A = - s \, f \,\ee^{4 \phi} \left(\frac12 g^{A{\bar B}} - \frac23 {\hat X}^A {\hat X}^B \right) Q_B \;, \label{ptx} 
\end{equation}
where we also employed \eqref{ginvdXdX} and \eqref{Gg}. 
Inserting \eqref{f2u} and \eqref{ptx} into \eqref{dtauXhorig} and using \eqref{relUnf} and \eqref{phiK} yields 
\begin{equation}
	\pd_\tau\Xh^A = \ee^U\left( 2 s \, \ee^{K/2}\Xh^A\Xh^B Q_B - \frac{1}{4} \ee^{-K/2}\Xh^A p^0 - s \, \ee^{K/2}g^{A\bar{B}}Q_B \right). 
	\label{dtauXh} 
\end{equation}
With the help of \eqref{lowerraiseA}, \eqref{Gg} and \eqref{phiK} we can express $\Xh^A$ as 
\begin{equation}
	\Xh^A = \ee^K g^{A\bar{B}} C_{BCD} \Xh^C \Xh^D \spc, 
\end{equation}
so that equation \eqref{dtauXh} becomes 
\begin{equation}
	\begin{split}
		\pd_\tau\Xh^A = \ee^U g^{A\bar{B}} &\left(2 s \, \ee^{3K/2}C_{BCD}\Xh^C\Xh^D Q_E\Xh^E - \frac{1}{4} \ee^{K/2}C_{BCD}\Xh^C\Xh^D p^0 - s \, \ee^{K/2} Q_B \right). 
		\label{dtauXhfinal} 
	\end{split}
\end{equation}
This expression precisely agrees with the flow equation for $\partial_{\tau} z^A$ given in \eqref{4dflow} and based on \eqref{W4}. 
Namely, evaluating 
\begin{equation}
	\begin{split}
		\pd_\tau z^A &=2 \ee^U g^{A\bar{B}}\pd_{\bar{B}} W_4 \\
		&= -\frac{1}{4}\ee^U g^{A\bar{B}}\pd_{\bar{B}}\left( \ee^{-K/2} p^0 + 8\ee^{K/2}\left| Q_A \Xh^A \right| \right) \\
		&= - \ee^U g^{A\bar{B}}\left[\left(\frac{1}{8}\, \ee^{K/2} p^0 - s \, \ee^{3K/2} Q_C \Xh^C\right)\pd_{\bar{B}}(\ee^{-K}) + \im s \, \ee^{K/2} Q_B\right] \\
		&= \im \ee^U g^{A\bar{B}}\left[\left(-\frac{1}{4}\, \ee^{K/2}p^0 + 2 s \, \ee^{3K/2} Q_E \Xh^E\right)C_{BCD} \Xh^C \Xh^D - s \, \ee^{K/2} Q_B\right] 
	\end{split}
	\raisetag{17.5ex} \label{dtauz} 
\end{equation}
shows that \eqref{dtauz} precisely equals \eqref{dtauXhfinal}. 
In deriving \eqref{dtauz} we used the relations 
\begin{equation}
	\pd_{\bar{B}} \left|Q_A \Xh^A \right| = s \, Q_A \pd_{\bar{B}}\Xh^A = \frac{\im}{2} s \, Q_B 
\end{equation}
and (from \eqref{emK}) 
\begin{equation}
	\pd_{\bar{A}}(\ee^{-K}) = -\frac{\im}{2} C_{ABC}(z^B-\bar{z}^B)(z^C-\bar{z}^C) = 2\im C_{ABC}\Xh^B\Xh^C \spc. 
\end{equation}

Next, we consider the $s=-1$ solution described in subsection \ref{floweqrot}. 
Proceeding as above, we compute 
\begin{eqnarray}
	\partial_{\tau} z^A &=& \partial_{\tau} \left( \alpha \, \ee^{-2 \phi} \,X^A \right)\nonumber\\
	&=& \left( - \frac23 \alpha f^2 N^{-1} {\rm e}^{- 2U} \left| W_5 \right| - p^0 \alpha f N^{-2} \ee^{-2U} \right. 
	\nonumber\\
	&& \left. 
	+ \frac{\im }{2} \Delta^{1/2} f N^{-1} \ee^{-U + 3 \phi} \left(p^0 \, \Delta \, \ee^{-6 \phi} - 2 Q_B {\hat X}^B \right) \right) X^A \nonumber\\
	&& + \alpha f \ee^{2 \phi} \left( \frac12 g^{A {\bar B}} - \frac23 {\hat X}^A {\hat X}^B \right) Q_B \;. 
\end{eqnarray}
Comparing with $\partial_{\bar B} W_4$ based on \eqref{w4zzbz}, 
\begin{eqnarray}
	\partial_{\bar B} W_4 &=& \im \ee^{3K/2} \Delta^{1/2} C_{BEF} {\hat X}^E {\hat X}^F \left( \frac18 p^0 \, \Delta \, \ee^{-K} - Q_A {\hat X}^A \right) \nonumber\\
	&& - \frac12 \alpha \, \ee^{K/2} \Delta^{-1/2} \left(\frac12 p^0 \, \Delta \, C_{BEF} {\hat X}^E {\hat X}^F - Q_B \right) \;, 
\end{eqnarray}
we find that \eqref{4dflow} precisely holds.

\bibliographystyle{JHEP-3} 
\bibliography{bibliography}

\providecommand{\href}[2]{#2}\begingroup\raggedright\begin{thebibliography}{10}

\bibitem{Ferrara:1995ih}
S.~Ferrara, R.~Kallosh and A.~Strominger, {\it {N = 2} extremal black holes},
  {\em Phys. Rev.} {\bf D52} (1995) 5412--5416
  [\href{http://arXiv.org/abs/hep-th/9508072}{{\tt hep-th/9508072}}].

\bibitem{Strominger:1996kf}
A.~Strominger, {\it Macroscopic entropy of {N = 2} extremal black holes},  {\em
  Phys. Lett.} {\bf B383} (1996) 39--43
  [\href{http://arXiv.org/abs/hep-th/9602111}{{\tt hep-th/9602111}}].

\bibitem{Ferrara:1996dd}
S.~Ferrara and R.~Kallosh, {\it Supersymmetry and attractors},  {\em Phys.
  Rev.} {\bf D54} (1996) 1514--1524
  [\href{http://arXiv.org/abs/hep-th/9602136}{{\tt hep-th/9602136}}].

\bibitem{Ferrara:1996um}
S.~Ferrara and R.~Kallosh, {\it Universality of supersymmetric attractors},
  {\em Phys. Rev.} {\bf D54} (1996) 1525--1534
  [\href{http://arXiv.org/abs/hep-th/9603090}{{\tt hep-th/9603090}}].

\bibitem{Ferrara:1997tw}
S.~Ferrara, G.~W. Gibbons and R.~Kallosh, {\it Black holes and critical points
  in moduli space},  {\em Nucl. Phys.} {\bf B500} (1997) 75--93
  [\href{http://arXiv.org/abs/hep-th/9702103}{{\tt hep-th/9702103}}].

\bibitem{Moore:1998pn}
G.~W. Moore, {\it Arithmetic and attractors},
  \href{http://arXiv.org/abs/hep-th/9807087}{{\tt hep-th/9807087}}.

\bibitem{Denef:2000nb}
F.~Denef, {\it Supergravity flows and {D}-brane stability},  {\em JHEP} {\bf
  08} (2000) 050 [\href{http://arXiv.org/abs/hep-th/0005049}{{\tt
  hep-th/0005049}}].

\bibitem{Behrndt:1996jn}
K.~Behrndt, G.~L. Cardoso, B.~de~Wit, R.~Kallosh, D.~L{\"u}st and T.~Mohaupt,
  {\it Classical and quantum {N = 2} supersymmetric black holes},  {\em Nucl.
  Phys.} {\bf B488} (1997) 236--260
  [\href{http://arXiv.org/abs/hep-th/9610105}{{\tt hep-th/9610105}}].

\bibitem{Sabra:1997kq}
W.~A. Sabra, {\it General static {N = 2} black holes},  {\em Mod. Phys. Lett.}
  {\bf A12} (1997) 2585--2590 [\href{http://arXiv.org/abs/hep-th/9703101}{{\tt
  hep-th/9703101}}].

\bibitem{Sabra:1997dh}
W.~A. Sabra, {\it Black holes in {N = 2} supergravity theories and harmonic
  functions},  {\em Nucl. Phys.} {\bf B510} (1998) 247--263
  [\href{http://arXiv.org/abs/hep-th/9704147}{{\tt hep-th/9704147}}].

\bibitem{Behrndt:1997ny}
K.~Behrndt, D.~L{\"u}st and W.~A. Sabra, {\it Stationary solutions of {N = 2}
  supergravity},  {\em Nucl. Phys.} {\bf B510} (1998) 264--288
  [\href{http://arXiv.org/abs/hep-th/9705169}{{\tt hep-th/9705169}}].

\bibitem{Sabra:1997yd}
W.~A. Sabra, {\it General {BPS} black holes in five dimensions},  {\em Mod.
  Phys. Lett.} {\bf A13} (1998) 239--251
  [\href{http://arXiv.org/abs/hep-th/9708103}{{\tt hep-th/9708103}}].

\bibitem{Chamseddine:1998yv}
A.~H. Chamseddine and W.~A. Sabra, {\it Metrics admitting {K}illing spinors in
  five dimensions},  {\em Phys. Lett.} {\bf B426} (1998) 36--42
  [\href{http://arXiv.org/abs/hep-th/9801161}{{\tt hep-th/9801161}}].

\bibitem{Gauntlett:1998fz}
J.~P. Gauntlett, R.~C. Myers and P.~K. Townsend, {\it Black holes of d = 5
  supergravity},  {\em Class. Quant. Grav.} {\bf 16} (1999) 1--21
  [\href{http://arXiv.org/abs/hep-th/9810204}{{\tt hep-th/9810204}}].

\bibitem{Gauntlett:2004qy}
J.~P. Gauntlett and J.~B. Gutowski, {\it General concentric black rings},  {\em
  Phys. Rev.} {\bf D71} (2005) 045002
  [\href{http://arXiv.org/abs/hep-th/0408122}{{\tt hep-th/0408122}}].

\bibitem{Gaiotto:2005gf}
D.~Gaiotto, A.~Strominger and X.~Yin, {\it New connections between 4d and 5d
  black holes},  {\em JHEP} {\bf 02} (2006) 024
  [\href{http://arXiv.org/abs/hep-th/0503217}{{\tt hep-th/0503217}}].

\bibitem{Gaiotto:2005xt}
D.~Gaiotto, A.~Strominger and X.~Yin, {\it 5d black rings and 4d black holes},
  {\em JHEP} {\bf 02} (2006) 023
  [\href{http://arXiv.org/abs/hep-th/0504126}{{\tt hep-th/0504126}}].

\bibitem{Gibbons:1997cc}
G.~W. Gibbons, {\it Supergravity vacua and solitons}, . Prepared for A Newton
  Institute Euroconference on Duality and Supersymmetric Theories, Cambridge,
  England, 7-18 Apr 1997.

\bibitem{Goldstein:2005hq}
K.~Goldstein, N.~Iizuka, R.~P. Jena and S.~P. Trivedi, {\it Non-supersymmetric
  attractors},  {\em Phys. Rev.} {\bf D72} (2005) 124021
  [\href{http://arXiv.org/abs/hep-th/0507096}{{\tt hep-th/0507096}}].

\bibitem{Tripathy:2005qp}
P.~K. Tripathy and S.~P. Trivedi, {\it Non-supersymmetric attractors in string
  theory},  {\em JHEP} {\bf 03} (2006) 022
  [\href{http://arXiv.org/abs/hep-th/0511117}{{\tt hep-th/0511117}}].

\bibitem{Kallosh:2006bt}
R.~Kallosh, N.~Sivanandam and M.~Soroush, {\it The non-{BPS} black hole
  attractor equation},  {\em JHEP} {\bf 03} (2006) 060
  [\href{http://arXiv.org/abs/hep-th/0602005}{{\tt hep-th/0602005}}].

\bibitem{Kallosh:2006ib}
R.~Kallosh, N.~Sivanandam and M.~Soroush, {\it Exact attractive non-{BPS} {STU}
  black holes},  {\em Phys. Rev.} {\bf D74} (2006) 065008
  [\href{http://arXiv.org/abs/hep-th/0606263}{{\tt hep-th/0606263}}].

\bibitem{Ceresole:2007wx}
A.~Ceresole and G.~Dall'Agata, {\it Flow equations for non-{BPS} extremal black
  holes},  {\em JHEP} {\bf 03} (2007) 110
  [\href{http://arXiv.org/abs/hep-th/0702088}{{\tt hep-th/0702088}}].

\bibitem{Andrianopoli:2007gt}
L.~Andrianopoli, R.~D'Auria, E.~Orazi and M.~Trigiante, {\it First order
  description of black holes in moduli space},
  \href{http://arXiv.org/abs/arXiv:0706.0712 [hep-th]}{{\tt arXiv:0706.0712
  [hep-th]}}.

\bibitem{Larsen:2006xm}
F.~Larsen, {\it The attractor mechanism in five dimensions},
  \href{http://arXiv.org/abs/hep-th/0608191}{{\tt hep-th/0608191}}.

\bibitem{Cardoso:2007rg}
G.~L. Cardoso, J.~M. Oberreuter and J.~Perz, {\it Entropy function for rotating
  extremal black holes in very special geometry},  {\em JHEP} {\bf 05} (2007)
  025 [\href{http://arXiv.org/abs/hep-th/0701176}{{\tt hep-th/0701176}}].

\bibitem{Nilsson:1984bj}
B.~E.~W. Nilsson and C.~N. Pope, {\it Hopf fibration of eleven-dimensional
  supergravity},  {\em Class. Quant. Grav.} {\bf 1} (1984) 499.

\bibitem{Duff:1997qz}
M.~J. Duff, H.~L{\"u} and C.~N. Pope, {\it Supersymmetry without
  supersymmetry},  {\em Phys. Lett.} {\bf B409} (1997) 136--144
  [\href{http://arXiv.org/abs/hep-th/9704186}{{\tt hep-th/9704186}}].

\bibitem{Behrndt:2005he}
K.~Behrndt, G.~L.~Cardoso and S.~Mahapatra, {\it Exploring the relation between
  4d and 5d {BPS} solutions},  {\em Nucl. Phys.} {\bf B732} (2006) 200--223
  [\href{http://arXiv.org/abs/hep-th/0506251}{{\tt hep-th/0506251}}].

\bibitem{Gunaydin:1983bi}
M.~G{\"u}naydin, G.~Sierra and P.~K. Townsend, {\it The geometry of {N = 2}
  {M}axwell-{E}instein supergravity and {J}ordan algebras},  {\em Nucl. Phys.}
  {\bf B242} (1984) 244.

\bibitem{Ferrara:2006xx}
S.~Ferrara and M.~G{\"u}naydin, {\it Orbits and attractors for {N = 2}
  {M}axwell-{E}instein supergravity theories in five dimensions},  {\em Nucl.
  Phys.} {\bf B759} (2006) 1--19
  [\href{http://arXiv.org/abs/hep-th/0606108}{{\tt hep-th/0606108}}].

\bibitem{Sahoo:2006rp}
B.~Sahoo and A.~Sen, {\it Higher derivative corrections to non-supersymmetric
  extremal black holes in {N = 2} supergravity},  {\em JHEP} {\bf 09} (2006)
  029 [\href{http://arXiv.org/abs/hep-th/0603149}{{\tt hep-th/0603149}}].

\bibitem{Cardoso:2006xz}
G.~L. Cardoso, B.~de~Wit and S.~Mahapatra, {\it Black hole entropy functions
  and attractor equations},  {\em JHEP} {\bf 03} (2007) 085
  [\href{http://arXiv.org/abs/hep-th/0612225}{{\tt hep-th/0612225}}].

\bibitem{Nampuri:2007gv}
S.~Nampuri, P.~K. Tripathy and S.~P. Trivedi, {\it On the stability of
  non-supersymmetric attractors in string theory},
  \href{http://arXiv.org/abs/arXiv:0705.4554 [hep-th]}{{\tt arXiv:0705.4554
  [hep-th]}}.

\bibitem{deWit:1996ix}
B.~de~Wit, {\it {N = 2} electric-magnetic duality in a chiral background},
  {\em Nucl. Phys. Proc. Suppl.} {\bf 49} (1996) 191--200
  [\href{http://arXiv.org/abs/hep-th/9602060}{{\tt hep-th/9602060}}].

\bibitem{Freedman:2003ax}
D.~Z. Freedman, C.~N{\'u}{\~n}ez, M.~Schnabl and K.~Skenderis, {\it Fake
  supergravity and domain wall stability},  {\em Phys. Rev.} {\bf D69} (2004)
  104027 [\href{http://arXiv.org/abs/hep-th/0312055}{{\tt hep-th/0312055}}].

\bibitem{Celi:2004st}
A.~Celi, A.~Ceresole, G.~Dall'Agata, A.~Van~Proeyen and M.~Zagermann, {\it On
  the fakeness of fake supergravity},  {\em Phys. Rev.} {\bf D71} (2005) 045009
  [\href{http://arXiv.org/abs/hep-th/0410126}{{\tt hep-th/0410126}}].

\bibitem{Ferrara:1997uz}
S.~Ferrara and M.~G{\"u}naydin, {\it Orbits of exceptional groups, duality and
  {BPS} states in string theory},  {\em Int. J. Mod. Phys.} {\bf A13} (1998)
  2075--2088 [\href{http://arXiv.org/abs/hep-th/9708025}{{\tt
  hep-th/9708025}}].

\bibitem{Bellucci:2006xz}
S.~Bellucci, S.~Ferrara, M.~G{\"u}naydin and A.~Marrani, {\it Charge orbits of
  symmetric special geometries and attractors},  {\em Int. J. Mod. Phys.} {\bf
  A21} (2006) 5043--5098 [\href{http://arXiv.org/abs/hep-th/0606209}{{\tt
  hep-th/0606209}}].

\bibitem{deWit:1984pk}
B.~de~Wit and A.~Van~Proeyen, {\it Potentials and symmetries of general gauged
  {N = 2} supergravity-{Y}ang-{M}ills models},  {\em Nucl. Phys.} {\bf B245}
  (1984) 89.

\bibitem{deWit:1984px}
B.~de~Wit, P.~G. Lauwers and A.~Van~Proeyen, {\it Lagrangians of {N = 2}
  supergravity-matter systems},  {\em Nucl. Phys.} {\bf B255} (1985) 569.

\end{thebibliography}\endgroup

\end{document}